\documentclass[times, twoside]{zHenriquesLab-StyleBioRxiv} 
\usepackage{blindtext}
\usepackage{lipsum}
\usepackage{amssymb}
\usepackage[utf8]{inputenc}
\usepackage{booktabs}   
\usepackage{multirow}   
\usepackage{array}

\usepackage{tabularx}
\usepackage{pifont}
\usepackage{wasysym}
\newcommand{\cmark}{\ding{51}} 
\newcommand{\xmark}{\ding{55}} 

\leadauthor{Mieko Ochi} 

\begingroup
\begin{document}

\title{Ensemble of Pathology Foundation Models for MIDOG 2025 Track 2: Atypical Mitosis Classification}
\shorttitle{Ensenble Learning for MIDOG 2025 Track2}

\author[1]{Mieko Ochi}
\author[1]{Yuan Bae}

\affil[1]{Department of Pathology, Japanese Red Cross Medical Center, Japan}

\maketitle

\begin{abstract}

Mitotic figures are classified into typical and atypical variants, with atypical counts correlating strongly with tumor aggressiveness. Accurate differentiation is therefore essential for patient prognostication and resource allocation, yet remains challenging even for expert pathologists. Here, we leveraged Pathology Foundation Models (PFMs) pre-trained on large histopathology datasets and applied parameter-efficient fine-tuning via low-rank adaptation. In addition, we incorporated ConvNeXt V2, a state-of-the-art convolutional neural network architecture, to complement PFMs. During training, we employed a fisheye transform to emphasize mitoses and Fourier Domain Adaptation using ImageNet target images. Finally, we ensembled multiple PFMs to integrate complementary morphological insights, achieving competitive balanced accuracy on the Preliminary Evaluation Phase dataset.

\end {abstract}


\begin{corrauthor}
noonecanpass.and@gmail.com, be\_yuan@med.jrc.or.jp
\end{corrauthor}

\section*{Introduction}

Hematoxylin and eosin–stained mitotic figure (MF) counts are essential for tumor evaluation, serving both as standalone and component grades in malignancy assessment \cite{intro1}. Mitotic figures are broadly classified into typical and atypical variants, with atypical forms—characterized by dysregulated chromatin aggregation and reflecting genomic instabilities such as chromosomal instability and aneuploidy—demonstrating independent prognostic value in cancers like breast carcinoma \cite{intro2,intro3}. However, manual enumeration and discrimination of MF variants are time-consuming and subject to substantial inter‐observer variability.

To address these challenges, we present a two-stage framework for automated MF classification in the MIDOG2025 Track 2 challenge \cite{ammeling_mitosis_2025}. First, we performed parameter-efficient fine-tuning of multiple Pathology Foundation Models (PFMs) using low-rank adaptation (LoRA) \cite{hu2021loralowrankadaptationlarge}. Training incorporated fisheye augmentation to emphasize central mitoses \cite{TOTH2022100339} and Fourier Domain Adaptation (FDA) for unsupervised style transfer with ImageNet images \cite{yang2021skunetmodelfourierdomain}. We further enhanced domain generalization by augmenting the MIDOG2025 set with an external labeled MF dataset \cite{jahanifar2025mitosis}. Second, we ensembled the adapted PFMs and ConvNeXt V2 \cite{woo2023convnextv2codesigningscaling} to integrate complementary morphological insights into a unified classification decision \cite{intro9}. Our method achieved a high balanced accuracy on validation splits and also demonstrated strong performance on the Preliminary Evaluation Phase dataset, underscoring its potential for reliable, automated MF analysis.


\section*{Material and Methods}

\subsection{Setting of the Training and Validation Datasets}

We trained all models using four publicly available datasets: the AMi-Br dataset \cite{bertram_histologic_2025}, the MIDOG 2025 Atypical Training Set \cite{weiss_2025_15188326}, the OMG-Octo Atypical dataset \cite{shen_2025_16107743}, and the dataset from Mostafa et al. \cite{jahanifar_2025_15390543}. The AMi-Br and MIDOG 2025 Atypical Training Set images were randomly split into training and validation subsets at a 4:1 ratio. The OMG-Octo Atypical and Mostafa et al.\ datasets were each used in their entirety for training.

In preprocessing, each input image was resized to 224 × 224 pixels with aspect ratio preserved and padded as necessary, and then underwent random brightness and contrast adjustments, followed by a random rotation. Next, an optical fisheye distortion was applied with the distortion coefficient sampled uniformly sampled from –0.9 to 0.9. FDA was applied with a probability of $p=0.5$, each time using a randomly selected target image drawn from a pool of 50,000 unlabeled ImageNet images.

For fine-tuning, we used the Adam optimizer (learning rate $1{\times}10^{-4}$, weight decay $1{\times}10^{-6}$ for ConvNeXt V2 and $1\times10^{-4}$ without weight deca for PFMs) and a batch size of 32 for all models. Training was performed for 50 epochs with early stopping. Additionally, we employed a \texttt{WeightedRandomSampler} with sampling weights of $1:0.15:0.15$ for the AMi-Br dataset + the MIDOG 2025 Atypical Training Set, OMG-Octo Atypical), and the Mostafa et al.\ dataset, respectively.

\subsection{LoRA for the Foundation Models and Ensemble Learning}

We fine-tuned ImageNet-pretrained ConvNeXt V2 (base, \texttt{fcmae\_ft\_in22k\_in1k}) with a binary classification head. For PFMs, based on the results reported by Banerjee et al. from LoRA fine-tuning  of foundation models on the AMi-Br dataset \cite{bertram_histologic_2025}, we selected three PFMs — UNI \cite{chen2024uni}, Virchow \cite{vorontsov2024virchowmillionslidedigitalpathology}, and Virchow2 \cite{zimmermann2024virchow2scalingselfsupervisedmixed} — for our own LoRA fine-tuning experiments.
Furthermore, following Haotian et al.’s method \cite{10981009} of applying LoRA to the query (Q) and value (V) projection matrices in the multi-head self-attention (MHSA) module, we introduced two low‐rank matrices $A_Q\in\mathbb R^{d\times r}$ and $B_Q\in\mathbb R^{r\times k}$, such that the update to the frozen query weight $W_Q\in\mathbb R^{d\times k}$ can be factorized as
\begin{align}
\Delta W_Q &= A_Q\,B_Q, \label{eq:lora-delta}\\
W_Q &= W_0 + \Delta W_Q
      = W_0 + A_Q\,B_Q\,. \label{eq:lora-final}
\end{align}
Only $A_Q$ and $B_Q$ are learned during fine‐tuning, while the original weight matrix $W_0$ remains fixed.

The foundation models’ divergent pretraining datasets led to substantial variations in feature extraction and diagnostic decision-making. To balance model diversity with clinical reliability, we developed a weighted ensemble framework that aggregates predictions from $N$ fine-tuned models.

Each base model $M_i$ produced a probability vector over the $C$ (=2) classes:
\begin{equation}
  P_i(x) = \bigl[p_i^{(1)}(x), \dots, p_i^{(C)}(x)\bigr], 
  \quad
  \sum_{c=1}^C p_i^{(c)}(x)=1.
\end{equation}

We then learn nonnegative weights $w_i$ on a validation set $\mathcal D_{\rm val}$ by maximizing diagnostic accuracy. Traditional overall accuracy maximization (Equation \ref{eq:balanced_ensemble}) may sacrifice performance on minority classes when class imbalance persists in the validation set. Therefore, we introduce a new objective function that directly maximizes balanced accuracy.
\begin{equation}
\label{eq:balanced_ensemble}
\begin{aligned}
  \mathbf{w}^* = \arg\max_{\mathbf{w}}\;&
    \frac{1}{C}\sum_{c=1}^{C} \frac{1}{|D_{\mathrm{val}}^{c}|}\\
  &\quad\times
    \sum_{(x,y)\in D_{\mathrm{val}}^{c}}
    \mathbf{1}\Bigl[y = \arg\max_{c'} \sum_{i=1}^{N} w_i\,p_i^{(c')}(x)\Bigr] \\
  \mathrm{s.t.}\;&
    w_i \ge 0, \quad \sum_{i=1}^{N} w_i = 1.
\end{aligned}
\end{equation}
Here, $C$ denotes the number of classes and $D_{\mathrm{val}}^{c}$ denotes the set of validation samples with label $c$. By equally weighting the accuracy of each class, we mitigate performance degradation on minority classes. Equation \ref{eq:balanced_ensemble} defines our proposed objective for the ensemble stage (i.e., maximizing balanced accuracy) under simplex constraints on the weights.
The ensemble weights were not updated jointly with base models. Instead, all individual models were fully trained first, and the weights were then optimized post hoc on the validation set using AutoGluon’s weighted ensemble (Caruana-style Ensemble Selection) \cite{erickson2020autogluontabularrobustaccurateautoml}.
Concretely, we employ a greedy forward selection with replacement over the candidate model library, iteratively adding the model that yields the largest increase in balanced accuracy on out-of-fold/validation predictions, and normalizing the nonnegative weights to sum to 1. This procedure is implemented via AutoGluon Tabular (v1.2) and corresponds to its default weighted-ensemble meta-learner.

Finally, the ensemble prediction is given by
\begin{equation}
  P_{\rm final}^{(c)}(x)
  = \sum_{i=1}^N w_i^*\,p_i^{(c)}(x),
  \quad
  \hat y = \arg\max_{c\in\{1,\dots,C\}}P_{\rm final}^{(c)}(x).
\end{equation}

\section*{Results}



As shown in Table 1, we conducted an ablation study using the UNI model on our validation dataset to assess the impact of fisheye and FDA augmentations during training. The highest balanced accuracy was achieved when fisheye augmentation was applied without FDA (85.457\%), followed by the combination of both fisheye and FDA (83.993\%).

\vspace{3mm}


\begin{table}[htbp]
  \centering
  \caption{Ablation of fisheye and FDA augmentation of UNI models on our validation dataset.}
  \label{tab:ablation-fisheye-fda-no-cat}
  \begin{tabular}{@{} c c c @{}}
    \toprule
     Fisheye & FDA  &Balanced Acc (\%) \\
    \midrule
     \xmark  & \xmark  &83.155\\
     \cmark  & \xmark  &85.457\\
     \xmark  & \cmark  &83.519\\
     \cmark  & \cmark  &83.993\\

    \bottomrule
  \end{tabular}
\end{table}

\vspace{3mm}
Furthermore, under the same augmentation settings, we compared the performance of each PFM and their ensemble on the validation dataset (Table 2). ConvNeXt V2 achieved the highest balanced accuracy among individual models, and ensembling all models yielded an improvement of approximately 10\% in balanced accuracy.

\vspace{3mm}
\begin{table}[htbp]
  \centering
  \caption{Performance comparison on our validation dataset.}
  \label{tab:balanced-acc-fisheye}
  \small 
  \begin{tabular}{@{} l p{4cm} c @{}}
    \toprule
    Category       & Model                     & Balanced Acc (\%) \\
    \midrule
    \multirow{4}{*}{Single Model}
                    & UNI          & 85.457                              \\
                   & Virchow     & 86.035                              \\
                   & Virchow2    & 87.590                              \\
                   & ConvNeXt V2    & 87.745                              \\
    \midrule
    \multirow{2}{*}{Ensemble Model}
                   & UNI + Virchow + Virchow2  & 93.568                             \\
                   & UNI + Virchow + Virchow2 + ConvNeXt V2  & 97.279                           \\
    \bottomrule
  \end{tabular}
\end{table}
\vspace{3mm}
Based on these results, in the Preliminary Evaluation Phase, we submitted ensemble models of PFMs trained under two conditions: fisheye only (Fisheye Only) and fisheye combined with FDA augmentation (Fisheye + FDA). As shown in Table 3, the overall balanced accuracy was higher with Fisheye + FDA, which consistently outperformed Fisheye Only across all domains. In other words, Fisheye + FDA exhibited stable performance, whereas Fisheye Only achieved strong results in some domains but showed greater variability overall. Finally, we evaluated an ensemble comprising all models, including ConvNeXt V2 trained with Fisheye + FDA augmentation, in the Preliminary Evaluation Phase. This ensemble achieved the highest overall balanced accuracy across all configurations.

\vspace{3mm}

\begin{table}[htbp]
\caption{Comparison of balanced accuracy (\%) between the submitted models across domains in the Preliminary Evaluation Phase. PFM denotes the ensemble of UNI, Virchow, and Virchow2, while ALL denotes the ensemble of UNI, Virchow, Virchow2, and ConvNeXt V2. OBA denotes overall balanced accuracy.}
\begin{tabularx}{\linewidth}{@{} XXXX @{}}
  \toprule
   & Fisheye Only (PFM) & Fisheye + FDA (PFM) & Fisheye + FDA (ALL) \\
  \midrule
  domain\_0 & 76.563 & 78.125 & 95.312 \\
  domain\_1 & 83.843 & 85.188 & 83.255 \\
  domain\_2 & 88.764 & 91.011 & 91.573 \\
  domain\_3 & 94.444 & 95.833 & 94.444 \\
  \midrule
  OBA & 86.803 & 88.371 & 88.879 \\
  \bottomrule
\end{tabularx}
\end{table}
\vspace{3mm}


\section*{Discussion}

We based our approach on the first‐place solution by Haotian et al. \cite{10981009} in the Pap Smear Cell Classification Challenge, combining parameter–efficient LoRA fine-tuning of individual PFMs with a subsequent ensemble learning stage. Banerjee et al. \cite{banerjee2025benchmarkingdeeplearningvision} demonstrated that, among the diverse PFMs evaluated on the AMi-Br dataset (one of the challenge’s benchmarks) \cite{bertram_histologic_2025}, Virchow2, UNI, and Virchow achieved the high balanced accuracies. Accordingly, we selected these three models for per-model adaptation. Each PFM has been pretrained on large collections of human histopathology images acquired under different conditions (institutions, countries, staining protocols, scanners, etc.), resulting in complementary attention patterns over cellular structures. By ensembling their predictions, we exploit these diverse inductive biases to arrive at more accurate final diagnoses. In addition, we incorporated ConvNeXt V2, a convolutional neural network (CNN) that achieved state-of-the-art performance across multiple vision benchmarks, including ImageNet recognition, in 2023, into the ensemble. Owing to its architectural design, CNNs extract features within local receptive fields of convolutional kernels, thereby capturing local patterns. In contrast, Vision Transformers extract features in parallel across the entire image, resulting in global receptive fields and holistic representations \cite{Geirhos2018ImageNettrainedCA,tuli2021convolutionalneuralnetworkstransformers}. Given that the target regions of interest in our task—mitotic figures—are confined to small local areas, we hypothesized that CNNs would provide superior accuracy.

Because distinguishing normal from atypical mitotic figures requires capturing subtle chromatin texture variations, we further augmented our training set with a fisheye transformation that emphasizes the central region of each image—where mitotic figures typically reside. Fisheye augmentation has previously been shown to boost single-cell classification accuracy in bladder, lung, and other tissues \cite{TOTH2022100339}. Moreover, to mitigate domain shifts arising from differences in scanner types and staining protocols, we applied style transfer via FDA using unlabeled natural images. Prior work by Yamashita et al. \cite{Yamashita_2021} and Yang et al.\cite{yang2021skunetmodelfourierdomain} has shown that natural image–based style transfer, including FDA, have been shown to substantially improve model performance across a variety of histopathological tasks—ranging from tumor gene subtyping to mitotic figure detection. Our results demonstrate the efficacy of this tailored adaptation strategy for histopathological image analysis and highlight its potential to reduce pathologists’ workload in clinical practice. In future work, we will explore advanced techniques to further enhance model generalization and validate our approach on larger, more diverse clinical datasets.


\thispagestyle{plain}
\section*{References}
\bibliography{literature}

\begin{thebibliography}{23}
\providecommand{\natexlab}[1]{#1}
\providecommand{\url}[1]{\texttt{#1}}
\expandafter\ifx\csname urlstyle\endcsname\relax
  \providecommand{\doi}[1]{doi: #1}\else
  \providecommand{\doi}{doi: \begingroup \urlstyle{rm}\Url}\fi

\bibitem[Donovan et~al.(2021)Donovan, Moore, Bertram, Luong, Bolfa,
  Klopfleisch, Tvedten, Salas, Whitley, Aubreville, and Meuten]{intro1}
Taryn~A. Donovan, Frances~M. Moore, Christof~A. Bertram, Richard Luong, Pompei
  Bolfa, Robert Klopfleisch, Harold Tvedten, Elisa~N. Salas, Derick~B. Whitley,
  Marc Aubreville, and Donald~J. Meuten.
\newblock Mitotic figures—normal, atypical, and imposters: A guide to
  identification.
\newblock \emph{Veterinary Pathology volume 58, issue 2, pages243–257}, 2021.

\bibitem[Ohashi et~al.(2018)Ohashi, Namimatsu, Sakatani, Naito, Takei, and
  Shimizu]{intro2}
Ryuji Ohashi, Shigeki Namimatsu, Takashi Sakatani, Zenya Naito, Hiroyuki Takei,
  and Akira Shimizu.
\newblock Prognostic utility of atypical mitoses in patients with breast
  cancer: A comparative study with ki67 and phosphohistone h3.
\newblock \emph{J Surg Oncol volume 118, issue 3, pages557-567}, 2018.

\bibitem[Lashen et~al.(2022)Lashen, Toss, Alsaleem, Green, and Rakha]{intro3}
Ayat Lashen, Michael~S. Toss, Mansour Alsaleem, Andrew~R Green, and Nigel P.
  Mongan \&~Emad Rakha.
\newblock The characteristics and clinical significance of atypical mitosis in
  breast cancer.
\newblock \emph{Modern Pathology volume 35, pages1341–1348}, 2022.

\bibitem[Ammeling et~al.(2025)Ammeling, Aubreville, Banerjee, Bertram,
  Breininger, Hirling, Horvath, Stathonikos, and Veta]{ammeling_mitosis_2025}
Jonas Ammeling, Marc Aubreville, Sweta Banerjee, Christof~A. Bertram, Katharina
  Breininger, Dominik Hirling, Peter Horvath, Nikolas Stathonikos, and Mitko
  Veta.
\newblock Mitosis {Domain} {Generalization} {Challenge} 2025.
\newblock Zenodo, March 2025.
\newblock \doi{10.5281/zenodo.15077361}.

\bibitem[Hu et~al.(2021)Hu, Shen, Wallis, Allen-Zhu, Li, Wang, Wang, and
  Chen]{hu2021loralowrankadaptationlarge}
Edward~J. Hu, Yelong Shen, Phillip Wallis, Zeyuan Allen-Zhu, Yuanzhi Li, Shean
  Wang, Lu~Wang, and Weizhu Chen.
\newblock Lora: Low-rank adaptation of large language models, 2021.

\bibitem[Toth et~al.(2022)Toth, Bauer, Sukosd, and Horvath]{TOTH2022100339}
Timea Toth, David Bauer, Farkas Sukosd, and Peter Horvath.
\newblock Fisheye transformation enhances deep-learning-based single-cell
  phenotyping by including cellular microenvironment.
\newblock \emph{Cell Reports Methods}, 2\penalty0 (12):\penalty0 100339, 2022.
\newblock ISSN 2667-2375.
\newblock \doi{https://doi.org/10.1016/j.crmeth.2022.100339}.

\bibitem[Yang et~al.(2021)Yang, Luo, Zhang, and
  Wang]{yang2021skunetmodelfourierdomain}
Sen Yang, Feng Luo, Jun Zhang, and Xiyue Wang.
\newblock Sk-unet model with fourier domain for mitosis detection, 2021.

\bibitem[Mostafa(2025)]{jahanifar2025mitosis}
Jahanifar Mostafa.
\newblock Mitosis subtyping dataset.
\newblock Zenodo, 2025.
\newblock [Data set].

\bibitem[Woo et~al.(2023)Woo, Debnath, Hu, Chen, Liu, Kweon, and
  Xie]{woo2023convnextv2codesigningscaling}
Sanghyun Woo, Shoubhik Debnath, Ronghang Hu, Xinlei Chen, Zhuang Liu, In~So
  Kweon, and Saining Xie.
\newblock Convnext v2: Co-designing and scaling convnets with masked
  autoencoders, 2023.

\bibitem[Dong et~al.(2020)Dong, Yu, Cao, Shi, and Ma]{intro9}
Xibin Dong, Zhiwen Yu, Wenming Cao, Yifan Shi, and Qianli Ma.
\newblock A survey on ensemble learning.
\newblock \emph{Frontiers of Computer Science volume 14, pages241–258}, 2020.

\bibitem[Bertram et~al.()Bertram, Weiss, Donovan, Banerjee, Conrad, Ammeling,
  Klopfleisch, Kaltenecker, and Aubreville]{bertram_histologic_2025}
Christof~A. Bertram, Viktoria Weiss, Taryn~A. Donovan, Sweta Banerjee, Thomas
  Conrad, Jonas Ammeling, Robert Klopfleisch, Christopher Kaltenecker, and Marc
  Aubreville.
\newblock Histologic dataset of normal and atypical mitotic figures on human
  breast cancer ({AMi}-br).
\newblock In Christoph Palm, Katharina Breininger, Thomas Deserno, Heinz
  Handels, Andreas Maier, Klaus~H. Maier-Hein, and Thomas~M. Tolxdorff,
  editors, \emph{Bildverarbeitung für die Medizin 2025}, pages 113--118.
  Springer Fachmedien Wiesbaden.
\newblock ISBN 978-3-658-47422-5.

\bibitem[Weiss et~al.(2025)Weiss, Banerjee, Donovan, Conrad, Klopfleisch,
  Ammeling, Kaltenecker, Hirling, Veta, Stathonikos, Horvath, Breininger,
  Aubreville, and Bertram]{weiss_2025_15188326}
Viktoria Weiss, Sweta Banerjee, Taryn Donovan, Thomas Conrad, Robert
  Klopfleisch, Jonas Ammeling, Christopher Kaltenecker, Dominik Hirling, Mitko
  Veta, Nikolas Stathonikos, Peter Horvath, Katharina Breininger, Marc
  Aubreville, and Christof Bertram.
\newblock A dataset of atypical vs normal mitoses classification for midog
  2025.
\newblock Zenodo, \url{https://doi.org/10.5281/zenodo.15188326}, 2025.

\bibitem[Shen et~al.(2025)Shen, Hawkins, Baer, Bräutigam, and
  Collins~Fekete]{shen_2025_16107743}
Zhuoyan Shen, Maria~Andreia Hawkins, Esther Baer, Konstantin Bräutigam, and
  Charles-Antoine Collins~Fekete.
\newblock Omg-octo atypical: A refinement of the original omg-octo database to
  incorporate atypical mitoses.
\newblock Zenodo, \url{https://doi.org/10.5281/zenodo.16107743}, 2025.

\bibitem[Jahanifar(2025)]{jahanifar_2025_15390543}
Mostafa Jahanifar.
\newblock Mitosis subtyping dataset.
\newblock Zenodo, \url{https://doi.org/10.5281/zenodo.15390543}, 2025.

\bibitem[Chen et~al.(2024)Chen, Ding, Lu, Williamson, Jaume, Chen, Zhang, Shao,
  Song, Shaban, et~al.]{chen2024uni}
Richard~J Chen, Tong Ding, Ming~Y Lu, Drew~FK Williamson, Guillaume Jaume,
  Bowen Chen, Andrew Zhang, Daniel Shao, Andrew~H Song, Muhammad Shaban, et~al.
\newblock Towards a general-purpose foundation model for computational
  pathology.
\newblock \emph{Nature Medicine}, 2024.

\bibitem[Vorontsov et~al.(2024)Vorontsov, Bozkurt, Casson, Shaikovski,
  Zelechowski, Liu, Severson, Zimmermann, Hall, Tenenholtz, Fusi, Mathieu, van
  Eck, Lee, Viret, Robert, Wang, Kunz, Lee, Bernhard, Godrich, Oakley, Millar,
  Hanna, Retamero, Moye, Yousfi, Kanan, Klimstra, Rothrock, and
  Fuchs]{vorontsov2024virchowmillionslidedigitalpathology}
Eugene Vorontsov, Alican Bozkurt, Adam Casson, George Shaikovski, Michal
  Zelechowski, Siqi Liu, Kristen Severson, Eric Zimmermann, James Hall, Neil
  Tenenholtz, Nicolo Fusi, Philippe Mathieu, Alexander van Eck, Donghun Lee,
  Julian Viret, Eric Robert, Yi~Kan Wang, Jeremy~D. Kunz, Matthew C.~H. Lee,
  Jan Bernhard, Ran~A. Godrich, Gerard Oakley, Ewan Millar, Matthew Hanna, Juan
  Retamero, William~A. Moye, Razik Yousfi, Christopher Kanan, David Klimstra,
  Brandon Rothrock, and Thomas~J. Fuchs.
\newblock Virchow: A million-slide digital pathology foundation model, 2024.

\bibitem[Zimmermann et~al.(2024)Zimmermann, Vorontsov, Viret, Casson,
  Zelechowski, Shaikovski, Tenenholtz, Hall, Klimstra, Yousfi, Fuchs, Fusi,
  Liu, and Severson]{zimmermann2024virchow2scalingselfsupervisedmixed}
Eric Zimmermann, Eugene Vorontsov, Julian Viret, Adam Casson, Michal
  Zelechowski, George Shaikovski, Neil Tenenholtz, James Hall, David Klimstra,
  Razik Yousfi, Thomas Fuchs, Nicolo Fusi, Siqi Liu, and Kristen Severson.
\newblock Virchow2: Scaling self-supervised mixed magnification models in
  pathology, 2024.

\bibitem[Jiang et~al.(2025)Jiang, Cai, Xu, Fang, Zhang, and Wang]{10981009}
Haotian Jiang, Jiangdong Cai, Mengjie Xu, Yu~Fang, Lichi Zhang, and Qian Wang.
\newblock Ensemble of foundation models for pap smear cell analysis.
\newblock In \emph{2025 IEEE 22nd International Symposium on Biomedical Imaging
  (ISBI)}, pages 1--4, 2025.
\newblock \doi{10.1109/ISBI60581.2025.10981009}.

\bibitem[Erickson et~al.(2020)Erickson, Mueller, Shirkov, Zhang, Larroy, Li,
  and Smola]{erickson2020autogluontabularrobustaccurateautoml}
Nick Erickson, Jonas Mueller, Alexander Shirkov, Hang Zhang, Pedro Larroy,
  Mu~Li, and Alexander Smola.
\newblock Autogluon-tabular: Robust and accurate automl for structured data,
  2020.

\bibitem[Banerjee et~al.(2025)Banerjee, Weiss, Donovan, Fick, Conrad, Ammeling,
  Porsche, Klopfleisch, Kaltenecker, Breininger, Aubreville, and
  Bertram]{banerjee2025benchmarkingdeeplearningvision}
Sweta Banerjee, Viktoria Weiss, Taryn~A. Donovan, Rutger H.~J. Fick, Thomas
  Conrad, Jonas Ammeling, Nils Porsche, Robert Klopfleisch, Christopher
  Kaltenecker, Katharina Breininger, Marc Aubreville, and Christof~A. Bertram.
\newblock Benchmarking deep learning and vision foundation models for atypical
  vs. normal mitosis classification with cross-dataset evaluation, 2025.

\bibitem[Geirhos et~al.(2018)Geirhos, Rubisch, Michaelis, Bethge, Wichmann, and
  Brendel]{Geirhos2018ImageNettrainedCA}
Robert Geirhos, Patricia Rubisch, Claudio Michaelis, Matthias Bethge, Felix
  Wichmann, and Wieland Brendel.
\newblock Imagenet-trained cnns are biased towards texture; increasing shape
  bias improve accuracy and robustness.
\newblock \emph{ArXiv}, abs/1811.12231, 2018.

\bibitem[Tuli et~al.(2021)Tuli, Dasgupta, Grant, and
  Griffiths]{tuli2021convolutionalneuralnetworkstransformers}
Shikhar Tuli, Ishita Dasgupta, Erin Grant, and Thomas~L. Griffiths.
\newblock Are convolutional neural networks or transformers more like human
  vision?, 2021.

\bibitem[Yamashita et~al.(2021)Yamashita, Long, Banda, Shen, and
  Rubin]{Yamashita_2021}
Rikiya Yamashita, Jin Long, Snikitha Banda, Jeanne Shen, and Daniel~L. Rubin.
\newblock Learning domain-agnostic visual representation for computational
  pathology using medically-irrelevant style transfer augmentation.
\newblock \emph{IEEE Transactions on Medical Imaging}, 40\penalty0
  (12):\penalty0 3945–3954, December 2021.
\newblock ISSN 1558-254X.
\newblock \doi{10.1109/tmi.2021.3101985}.

\end{thebibliography}

\end{document}